\documentclass[twocolumn ,amssymb, amsmath,nobibnotes, aps, prl]{revtex4}

\expandafter\ifx\csname package@font\endcsname\relax\else
 \expandafter\expandafter
 \expandafter\usepackage
 \expandafter\expandafter
 \expandafter{\csname package@font\endcsname}%
\fi

\begin{document}

\title{Weinberg-Salam model at finite temperature and density}
\author{M. Loewe}
\email{mloewe@fis.puc.cl}
\affiliation{Facultad de F\'\i
sica, Pontificia Universidad Cat\'olica de Chile,\\ Casilla 306,
Santiago 22, Chile}
\author{S. Mendizabal}
\email{smendizabal@fis.puc.cl}
\affiliation{Facultad de F\'\i
sica, Pontificia Universidad Cat\'olica de Chile,\\ Casilla 306,
Santiago 22, Chile}
\author{J. C. Rojas}
\email{jcristobalrojas@hotmail.com} \affiliation{Departamento de
F\'\i sica, Universidad Cat\'olica del Norte,\\ Casilla 1280,
Antofagasta, Chile.}

\begin{abstract}

 We present a new gauge fixing condition for the Weinberg-Salam electro-weak theory at finite temperature and density.
  After spontaneous symmetry breaking occurs, every unphysical term in the Lagrangian is eliminated with our gauge
  fixing condition. A new and simple Lagrangian can be obtained where we can identify the propagators and vertices.
   Some consequences are discussed, as the new gauge dependent
masses of the gauge fields and the new
  Faddeev-Popov Lagrangian. After obtaining
the quadratic terms, we calculate exactly the 1-loop effective
potential identifying the contribution
 of every particular field.
\end{abstract}

\maketitle

The possibility of a new mechanism leading to spontaneous symmetry
breaking induced through Bose-Einstein condensation, due to
chemical potentials, has recently been explored in the frame of
the electro-weak model \cite{Sannino1}. There is a crucial
difference with the standard symmetry breaking mechanism since now
the number of Nambu-Goldstone bosons that appear is lesser than
the number required by the Goldstone theorem \cite{miransky}. This
is a new possibility for discussing, in general, phase transitions
in field theory.

 This idea has been implemented, for example, in a scalar model with a chemical
potential associated to the total conserved charge
\cite{kapusta1,haber1,haber2,actor}. In this model, the exact
calculation of the effective
 potential has been achieved \cite{bernstein,benson}. Another
 example of condensation is the occurrence of a condensed phase for
 a pion system
 when the isospin chemical potential becomes bigger then a critical
 value dependent on temperature \cite{LV}.

On the other hand, the existence of a gauge field condensation
\cite{linde}, has motivated many authors
 to consider the existence of chemical potentials associated to those fields, allowing the
  introduction of a conserved charges in gauge invariant theories. In this way
   $W$ condensation \cite{chaichian,kapusta2}
  or vector meson condensation \cite{sannino2} can be induced. The issue of relativistic Bose condensed vector fields in
  strong interactions was
  suggested for the first time in \cite{sannino3}.This fact is related to a spontaneous breaking of the
  rotational symmetry  when $\mu^{2}>m^{2}$, as seen in \cite{miransky} and
  \cite{gusynin}. The classification of all possible cases of
  rotational symmetry breaking due to a massive relativistic
  vector condensation was first considered in \cite{sannino4}.

Gauge fixing conditions in scenarios with finite temperature and
density have been matter of discussion for a long time. It
 is well known that some gauges, as for example the unitary gauge, do not give  precise
 values for the transition temperature. We  expect that 't Hooft $R_{\xi}$-type
 gauges
  would be more natural to discuss these kind of phenomena.
  Gauge invariant theories with finite temperature and density
   are difficult to deal with, since it is not easy to identify the propagators of the matter fields
   due to the mixing with the gauge fields. Therefore, the exact calculation of the effective potential is impossible without
   the use of tools like the high temperature expansion.

In this letter, we address how to manage the Weinberg-Salam model
in the presence of an $SU(2)$ and $U(1)$ chemical potential. The
thermodynamics of the standard model, including the phase
structure, has been discussed by Gynter in the frame of
dimensional reduction \cite{Gynter}. We will consider the model
under the perspective of a new gauge fixing condition, which
enable to remove all unphysical terms of the Lagrangian. This
allows us to compute the effective potential exactly. A high
temperature expansion is then considered in order to compare with
previous results, giving support to our treatment.


Let us consider the Weinberg-Salam model
\begin{eqnarray}
\mathcal{L}&=& -\frac{1}{4}F^{a}_{\mu\nu}F^{\mu\nu}_{a}-\frac{1}{4}B_{\mu\nu}B^{\mu\nu}+(D_{\mu}\phi)^{\dagger}(D^{\mu}\phi)\nonumber\\
&&+\bar{\psi}_{L}D\hspace{-0.25 cm}/\psi_{L}+\bar{e}_{R}(i\partial_{\mu}\hspace{-0.35 cm}/\hspace{0.2 cm}-g'B_{\mu}\hspace{-0.4 cm}/\hspace{0.2 cm})e_{R}\nonumber\\
& & -G(\bar{\psi}_{L}\phi e_{R}+\bar{e}_{R}\phi^{\dagger}\psi_{L})-V(\phi),
\end{eqnarray}
\noindent where
\begin{eqnarray}
D_{\mu}\phi&=&\partial_{\mu}-\frac{1}{2}igA_{\mu}^{a}\tau^{a}-\frac{1}{2}ig'B_{\mu},\nonumber\\
D_{\mu}{\psi}_{L}&=&\partial_{\mu}+\frac{1}{2}gA_{\mu}^{a}\tau^{a}-\frac{1}{2}g'B_{\mu}.
\end{eqnarray}
\noindent $\psi_{L}$ and $e_{R}$ represent the doublet and singlet leptons respectively, defined by
\begin{eqnarray}
\psi_{L}&=&\left[\begin{array}{c}v_{e},\\
e^{-}\end{array}\right]_{L},\\
e_{L}^{-}&=&\frac{1}{2}(1-\gamma_{5})e^{-},\\
e_{R}&=&\frac{1}{2}(1+\gamma_{5})e^{-},
\end{eqnarray}
\noindent and the scalar doulet is given by
\begin{eqnarray}
\phi=\left(\begin{array}{c}G^{+}\\ \frac{1}{\sqrt{2}}(H+iG^{0})\end{array}\right),
\end{eqnarray}
\noindent where $G^{+}$, $G^{0}$ y $H$ are the charged, neutral
and Higgs boson, respectively. The classical potential is given by
$V(\phi)=m^{2}\phi^{\dagger}\phi+\frac{1}{4}\lambda(\phi^{\dagger}\phi)^{2}$.

This Lagrangian is invariant under  $SU(2)\times U(1)$
transformation. The introduction of chemical potentials is
performed in the usual way. This model exhibits five conserved
charges: hypercharge associated to $U(1)$, Isospin charges from
$SU(2)$ and, finally, the leptonic charge. It is better to work
with a linear combination of those charges, leaving us with a set
of two new conserved commuting charges, plus the leptonic charge
$Q_{lep}$.
\begin{eqnarray}
Q_{el} &=& Q_{U(1)}+Q^{3}_{SU(2)},\\ \nonumber Q_{W} &=&
\frac{2}{\cos{2\theta_{w}}}(\sin^{2}{\theta_{w}}Q_{U(1)}-\cos^{2}{\theta_{w}}Q^{3}_{SU(2)}).
\end{eqnarray}
The charges $Q_{el}$ and $Q_{W}$ are the electromagnetic and
neutral-weak charges.
 $\theta_{w}$ is the Weinberg angle. Each chemical potential is introduced in the partition function as a
 Lagrange multiplier
\begin{equation}
Z=Tr\exp{[-\beta(H-\mu_{1}Q_{el}-\mu_{2}Q_{W}-\mu_{3}Q_{lep})]}.
\end{equation}
After identifying and integrating the conjugate momenta, we obtain
the following expression for the partition function
\begin{equation}\label{partition}
Z=\int{[d\Psi_{i}]\exp{\int_{0}^{\beta}{\int{d^{3}x[\tilde{\mathcal{L}}+\mu_{3}Q_{lep}]}}}}.
\end{equation}
Notice that we are integrating over each field. The modified
Lagrangian $\tilde{\mathcal{L}}$ is the old one but with the
replacement
\begin{eqnarray}
B_{\mu} &\rightarrow& B_{\mu}-\left(\mu_{1}+\frac{2\sin^{2}{\theta_{w}}}{\cos{2\theta_{w}}}\mu_{2}\right)\frac{1}{g'}v_{\mu}, \nonumber\\
A_{\mu}^{a} &\rightarrow& A_{\mu}^{a}-\left(\mu_{1}-\frac{2\cos^{2}{\theta_{w}}}{\cos{2\theta_{w}}}\mu_{2}\right)\frac{1}{g}\delta^{a3}v_{\mu},\label{sust}
\end{eqnarray}
\noindent where $v_{\mu}$ is a 4-velocity with respect to the thermal vacuum.\\
Before considering spontaneous symmetry breaking, let us express
our Lagrangian in terms of the shifted fields.
\begin{eqnarray}
A_{\mu} & = & B_{\mu}\cos{\theta}+A^{3}_{\mu}\sin{\theta},\nonumber\\
Z_{\mu} & = & B_{\mu}\sin{\theta}-A^{3}_{\mu}\cos{\theta},\nonumber\\
W^{\pm}_{\mu} & = & \frac{1}{\sqrt{2}}(W^{1}_{\mu} \mp
W^{2}_{\mu}).
\end{eqnarray}
\noindent In this way, we can rewrite every partial derivative of
a particular field with an associate chemical
 potential: $\partial^{j}_{\mu}\rightarrow\partial^{j}_{\mu}-i\mu^{j}v_{\mu}$, where the index $j$
 represents
  each particular field. So we define the following chemical potentials:\\
\noindent Scalar bosons
\begin{eqnarray}
G^{\pm} & \rightarrow & \pm \mu_{G}\equiv \mu_{1}-\mu_{2},\\
H^{\pm}=\frac{1}{\sqrt{2}}(H\pm iG^{0}) & \rightarrow &
\pm\mu_{H}\equiv \frac{\mu_{2}}{\cos{2\theta}}.
\end{eqnarray}
\noindent Gauge fields:
\begin{eqnarray}
W^{\pm}_{\mu} & \rightarrow & \pm \mu_{W}\equiv
\mu_{1}-\frac{2\cos{^{2}\theta}}{\cos{2\theta}}\mu_{2}.
\end{eqnarray}
\noindent Notice that there are no chemical potentials associated
to the $\gamma$ and $Z^{0}$. \noindent Leptons
\begin{eqnarray}
e_{R} & \rightarrow & \mu_{e_{R}}=\mu_{1}+\frac{2\sin{^{2}\theta}}{\cos{2\theta}}\mu_{2}+\mu_{3},\\
e_{L} & \rightarrow & \mu_{e_{L}}=\mu_{1}-\mu_{2}+\mu_{3},\\
\nu & \rightarrow & \mu_{\nu}=\frac{\mu_{2}}{cos{2\theta}}+\mu_{3}.
\end{eqnarray}
Gauge boson
 condensation occurs  when $\mu_{W}^{2}>m_{W}^{2}$ which is
 equivalent to $(\mu_{e_L} - \mu_{\nu})^2 > m_{W}^{2}$, i.e. when leptonic matter becomes extremely asymmetric.
 Since $\mu _{3}Q_{lep}$ appears as an additive term in (\ref{partition}), we will not consider this it in the gauge fixing conditions
 that follows. The replacement (\ref{sust}) will give new masses to the scalar and
  gauge fields, producing new propagators and vertices. Nevertheless,  the major conflict
   we must solve is how to handle the mixing terms between the scalar and gauge fields.


When a spontaneous symmetry breaking occurs, new features must be
taken in advance. As usual we expand the scalar field around a new
vacuum. We take
\begin{eqnarray}
\phi=\frac{1}{\sqrt{2}}\left(\begin{array}{c}0\\ \nu\end{array}\right)+\left(\begin{array}{c}G^{+}\\ \frac{1}{\sqrt{2}}(H+iG^{0})\end{array}\right),
\end{eqnarray}
The best way to treat the new mixing quadratic terms that appear
in the model is through an adequate gauge fixing condition. The
gauge we propose belongs to  the $R_{\xi}$-type gauges used by 't
Hooft, in which every quadratic
 mixed term disappears. However, we pay the price of having new gauge dependent
 masses. Nevertheless, the Faddeev-Popov Lagrangian
 will ensure that every gauge dependent term will not appear as a physical quantity.\\

The
 new gauge fixing condition is the following. For $U(1)$:
\begin{eqnarray}
F &\equiv&
(\partial_{\mu}-iC^{3}v_{\mu})B_{\mu}+ig'\xi\phi^{\dagger}<\phi>,
\end{eqnarray}
\noindent and for $SU(2)$
\begin{equation}
F^{a} \equiv
(\partial_{\mu}-iC^{a}v_{\mu})A^{a}_{\mu}+ig\xi\phi^{\dagger}\tau^{a}<\phi>,
\end{equation}
\noindent for each $a=1,2,3$. Here we have defined
\begin{eqnarray}
C^{3}& \equiv& \frac{2}{\cos{2\theta_{w}}}\mu_{2},\nonumber\\
 C^{1}=C^{2}& \equiv & \mu_{1}+\frac{2\sin^{2}{\theta_{w}}}{\cos{2\theta_{w}}}\mu_{2},
\end{eqnarray}
\noindent These functions enter in the Lagrangian in the following way
\begin{equation}
\mathcal{L}_{GF}=-\frac{1}{2\xi}|F|^2-\frac{1}{2\xi}|F^{a}|^{2}.
\end{equation}
Although this new gauge fixing condition assures us that we can
obtain a simple model, where we can properly identify quadratic
terms for each field and of course the propagators, the gauge
fields masses should be treated carefully.

Our gauge fixing condition facilitates the determination of the
1-loop effective potential, compared with the usual procedure,
since we are able to calculate the quadratic terms of every single
field in the model. After performing the sums over Matsubara
frequencies, the contribution of the fields has the form
\begin{widetext}
\begin{eqnarray}
\Omega_{eff}^{1-loop}= \Omega^0 +
\Omega_{G^{\pm}}+\Omega_{H,G^{0}}+\Omega_{\gamma}
+\Omega_{Z}+\Omega_{W^{\pm}}+\Omega_{ghost},
\end{eqnarray}
\noindent where $\Omega^0$ denotes the tree level term and each
one-loop term is given by
\begin{eqnarray}\label{potefectivo}
\Omega^{i}_{eff}=\int{\frac{d^{3}k}{(2\pi)^{3}}\left[\frac{a_{i}E^{+}_{i}+b_{i}E^{-}_{i}}{2}
+\frac{1}{\beta}\left( a_{i}\ln{(1-e^{-\beta E^{+}_{i}})}+b_{i}\ln{(1-e^{-\beta E^{-}_{i}})}\right)\right]}.
\end{eqnarray}
\end{widetext}
\noindent The subindex $i$ in the previous equation  refers to
each contribution. $a_{i}$ and $b_{i}$
 are constants related to $E^{\pm}_{i}$. These constants are associated to the number of degrees of freedom of each
 field.
As expected, there are no new divergencies associated  to
temperature and chemical potential, so the renormalizability is
maintained. We took $\xi=1$ to simplify the calculation. However,
it is possible to keep track of the gauge dependent parameters. In
any case, those terms vanish in our final result for the effective
potential when the ghost terms are properly taken into account.
 Referring to equation (\ref{potefectivo}), the energy
spectra for the different fields are: scalar fields
 $G^{\pm}$ where $a_{1}=b_{1}=1$
\begin{eqnarray}
E^{\pm}_{1}=\sqrt{\textbf{k}^{2}+m^{2}_{G}}\mp\mu_{G},
\end{eqnarray}
\noindent for $H$ and $G^{0}$, where $a_{2}=b_{2}=1$
\begin{eqnarray}\label{energias}
E^{\pm}_{2}=\left[\frac{A\mp\sqrt{A^{2}-4B}}{2}\right]^{1/2},
\end{eqnarray}
\noindent where
\begin{eqnarray}
A=2\textbf{k}^{2}+m^{2}_{H}+m^{2}_{G^{0}}+2\mu^{2}_{H},
\nonumber
\\
B=(\textbf{k}^{2}+m^{2}_{H}-\mu^{2}_{H})(\textbf{k}^{2}+m^{2}_{G^{0}}-\mu^{2}_{H}),
\end{eqnarray}
\noindent being $m_{G}$, $m_{G^{0}}$ and  $m_{H}$, mass parameters
associated to the $G^{\pm}$, $H$ and $G^{0}$ fields, respectively.
\begin{eqnarray}
m_{G}^{2} & = & m^{2}+\frac{1}{4}\lambda\nu^{2}+m_{W}^{2},\nonumber\\
m_{G^{0}}^{2} & = & m^{2}+\frac{1}{4}\lambda\nu^{2}+m_{Z}^{2},\nonumber\\
m_{H}^{2} & = & m^{2}+\frac{3}{4}\lambda\nu^{2},\nonumber\\
\end{eqnarray}
with
\begin{equation}
m_{Z}^{2}=\frac{(g'^{2}+g^{2})}{4}\nu^{2} \qquad
m_{W}^{2}=\frac{g^{2}\nu^{2}}{4}
\end{equation}
Here we included
the tree level mass term $m$, the mass due to the vacuum
expectation value and the gauge dependent mass term associated
to the gauge fixing condition. Due to the Goldstone theorem \cite{love},
the Higgs field has no gauge dependent mass contribution. \\
\noindent Now the excitation energies for the photon $\gamma$,
where $a_{3}  =  1$, $b_{3}  = 3$, are
\begin{eqnarray}
E^{+}_{3}=\sqrt{\textbf{k}^{2}+\tilde m_{\gamma}^{2}}, \qquad E^{-}_{3}=|\textbf{k}|.
\end{eqnarray}
\noindent For the $Z^{0}$, where $a_{4} = 1$, $b_{4} = 3$
\begin{eqnarray}
E^{+}_{4}=\sqrt{\textbf{k}^{2}+\tilde{m}^{2}_{Z}},\qquad
E^{-}_{4}=\sqrt{\textbf{k}^{2}+m^{2}_{Z}}.
\end{eqnarray}
\noindent The $W^{\pm}$ bosons are a little more complicated. The
result can be separated into two parts, which means $a_{5}=b_{5}=1$, $a_{6}=b_{6}= 3$
\begin{eqnarray}
E^{\pm}_{5} &=& \sqrt{\textbf{k}^{2}+\tilde m^{2}_{W}+2\mu_{W}^{2}},\\
E^{\pm}_{6}&=&\sqrt{\textbf{k}^{2}+m^{2}_{W}}\mp\mu_{W}.
\end{eqnarray}
with
\begin{equation}
\tilde{m}_{\gamma}=\tilde{m}_{Z}=C^{3}\qquad\tilde{m}_{W}=C^{1}
\end{equation}
\noindent Notice that $\tilde m_{\gamma}$, $\tilde{m}_{Z}$ and
$\tilde{m}_{W}$, in the previous equations, are gauge dependent
masses. They only appear in the extra spurious degrees of freedom
of the effective potential that normally are found  in finite
temperature calculations \cite{Bernard}. Finally, they will be
cancelled with the aid of the Faddeev-Poppov ghosts

\begin{eqnarray}
\mathcal{L}_{F-P}= &-& \bar{\eta}(\partial_{\mu}\partial^{\mu}+iC^3 v_{\mu}\partial^{\mu})\eta\nonumber\\
&-&\xi g'^{2}\bar{\eta}(\phi^{\dagger}+<\phi>^{\dagger})<\phi>\eta\nonumber\\
&-& \bar{\eta}^{a}(\partial_{\mu}+iC^{a}v_{\mu})(\partial^{\mu}\eta^{a}+g\epsilon^{abc}A_{\mu}^{c}\eta^{b})\nonumber\\
&-&
2g^{2}\xi\bar{\eta}^{a}(\phi^{\dagger}+<\phi>^{\dagger})\tau^{b}\tau^{a}<\phi>\eta^{b},
\end{eqnarray}

The energy spectra for the ghosts have the same coefficients,
$a_{7}=b_{7}=a_{8}=b_{8}=a_{9}=b_{9}=a_{10}=b_{10}=-1$.
$E_{7}^{\pm}$ and $E_{8}^{\pm}$ have the same form as
(\ref{energias}) but with different $A$ and $B$. For $E_{7}^{\pm}$

\begin{eqnarray}
A_7 &=& 2(\textbf{k} ^{2}+\frac{g'^{2}\nu^{2}}{4})+(C^{3})^{2},\nonumber\\
B_7 &=& (\textbf{k} ^{2}+\frac{g'^{2}\nu^{2}}{4})^{2}.
\end{eqnarray}

For $E_{8}^{\pm}$

\begin{eqnarray}
A_8 &=& 2(\textbf{k} ^{2}+\frac{g^{2}\nu^{2}}{4})+(C^{3})^{2},\nonumber\\
B_8 &=& (\textbf{k} ^{2}+\frac{g^{2}\nu^{2}}{4})^{2}.
\end{eqnarray}
the remaining ghost contribution differ a little from the above.
The result is given by
\begin{eqnarray}
E^{\pm2}_{9,10}=\frac{\hat{A}\mp\sqrt{\hat{A}^{2}-4\hat{B}}}{2},
\end{eqnarray}
for $E_{9}^{\pm}$
\begin{eqnarray}
\hat{A}_9 &=& 2(\textbf{k} ^{2}+\frac{g^{2}\nu^{2}}{2})+(C^{3})^{2},\nonumber\\
\hat{B}_9 &=& (\textbf{k} ^{2}+\frac{g^{2}\nu^{2}}{2})^{2}.
\end{eqnarray}
And $E_{10}^{\pm}$
\begin{eqnarray}
\hat{A}_{10} &=& 2\textbf{k} ^{2}+(C^{3})^{2},\nonumber\\
\hat{B}_{10} &=& \textbf{k} ^{4}.
\end{eqnarray}

These results correspond to an exact calculation of the effective
potential. The high temperature expansion gives us the same gauge
independent result obtained by Kapusta in \cite{kapusta2}

\begin{eqnarray}
\Omega_{eff}^{T} &=& -12\frac{\pi^{2}}{90}T^{4}+\frac{T^{2}}{24}[4m^{2}-4\mu_{G}^{2}-4\mu_{H}^{2}+\frac{6}{4}\lambda\nu^{2}]\nonumber\\
&&+\frac{T^{2}}{24}[3(2m_{W}^{2}+m_{Z}^{2})-8\mu_{W}^{2}]\nonumber\\
&&+\frac{\lambda\nu^{4}}{16}+\frac{\nu^{2}}{2}(m^{2}-\mu_{H}^{2}).
\end{eqnarray}

In the previous expression, we have not shown the contribution
from the leptons as well as the quartic terms in the chemical
potentials.

We would like to remark that our gauge fixing condition simplifies
considerably the procedure of calculating the effective potential,
since every single term can be identified unambiguously, avoiding
mixtures.

\begin{acknowledgments}
M.L. and S.M. would like to thank Professor 't Hooft for very
helpful remarks concerning our gauge fixing condition. Support
from FONDECYT 1010976 is acknowledged.
\end{acknowledgments}


\begin{thebibliography}{8.}

\bibitem{Sannino1} F. Sannino, K. Tuominen: Phys: Rev. D \textbf{68}, 016007
(2003)

\bibitem{miransky} V. A. Miransky, I. A. Shovkovy: Mod. Phys. Lett. A \textbf{19}, 1341
(2004)

\bibitem{kapusta1} J. I. Kapusta: Phys: Rev. D \textbf{24}, 426
(1981)

\bibitem{haber1} H. E. Haber, H. A. Weldon: Phys. Rev. Lett. \textbf{46}, 1497
(1981)

\bibitem{haber2} H. E. Haber, H. A. Weldon: Phys. Rev. D \textbf{25}, 502
(1982)

\bibitem{actor} A. Actor: Phys. Lett. B \textbf{157}, 53
(1985)

\bibitem{bernstein} J. Bernstein, S. Dodelson: Phys. Rev. Lett. \textbf{66}, 683
(1991)

\bibitem{benson} K. M. Benson, J. Bernstein, S. Dodelson: Phys. Rev. D \textbf{44},
2480 (1991)

\bibitem{LV} M. Loewe and C. Villavicencio: Phys. Rev. D
\textbf{67}, 074034 (2003), {\it ibid.} Phys. Rev. D \textbf{70},
074005 (2004)

\bibitem{linde} A. D. Linde: Phys. Lett. B \textbf{86}, 39
(1979)

\bibitem{chaichian} M. Chaichian, R. Gonzalez Felipe, D. Louis Martinez, H. Perez Rojas: CERN-TH-5870-90
(1990)

\bibitem{kapusta2} J. I. Kapusta: Phys. Rev. D \textbf{42}, 919
(1990)

\bibitem{sannino2} F. Sannino: hep-ph/0307053
(2003)

\bibitem{sannino3} J. T. Lenighan, F. Sannino, K. Splittorff: Phys
Rev. D \textbf{65}, 054002 (2002)

\bibitem{sannino4} F. Sannino: Phys
Rev. D \textbf{67}, 054006 (2003)

\bibitem{gusynin} V. P. Gusynin, V. A. Miransky, I. A. Shovkovy: Mod. Phys. Lett. A \textbf{19}, 1341
(2004)

\bibitem{Gynter} A. Gynther: Phys. Rev. D \textbf{68}, 016001
(2003)

\bibitem{love} D. Bailin, A. Love: {\it Introduction to Gauge Field
Theory}, (Adam Hilger, 1986), pp. 169-190

\bibitem{Bernard} C. W. Bernard: Phys. Rev. D \textbf{9}, 3312
(1974)

\end{thebibliography}
\end{document}